\documentclass[usenatbib]{mn2e}
\usepackage{graphicx}
\usepackage{pslatex}
\usepackage{float}
\usepackage{wasysym}

\usepackage[]{hyperref}

\usepackage{color}
\definecolor{mygreen}{rgb}{0.0,0.6,0}
\definecolor{myblue}{rgb}{0.2,0.0,0.7}
\definecolor{mybrown}{rgb}{0.5,0.2,0.0}

\newcommand\hide[1]{}
\author[B. Gauza et al.]{
\hspace*{-2.3mm} B. Gauza$^{1,2}$~\thanks{e-mail:bgauza@iac.es}, 
V.~J.~S. B\'ejar$^{1,2}$, 
R. Rebolo$^{1,2,3}$,
C. \'Alvarez$^{1,2}$, 
G. Bihain$^{4}$,
\newauthor M. R. Zapatero Osorio$^{5}$,
~J. A. Caballero$^{5}$, 
~C. M. Telesco$^{6}$, 
~C. Packham$^{7}$ 
\newauthor 
\\
\\
$^1$Instituto de Astrof\'{i}sica de Canarias (IAC), E-38200 La Laguna, Tenerife, Spain\\
$^2$Dept. Astrof\'{i}sica, Universidad de La Laguna (ULL), E-38206 La Laguna, Tenerife, Spain\\
$^3$Consejo Superior de Investigaciones Cient\'ificas, CSIC, E-28006 Madrid, Spain \\
$^4$Leibniz-Institut f\"{u}r Astrophysik Potsdam (AIP), An der Sternwarte 16, D-14482 Potsdam, Germany\\
$^5$Centro de Astrobiolog\'ia (CSIC-INTA), Ctra. Ajalvir km 4, E-28850, Torrej\'on de Ardoz, Madrid, Spain\\
$^6$Department of Astronomy, University of Florida, Gainesville, FL 32611, USA\\
$^7$Department of Physics and Astronomy, University of Texas at San Antonio, San Antonio, TX 78249, USA\\
}
\begin{document}
%
%
\title[CanariCam mid-IR imaging of the Barnard's Star]{Constraints on the substellar companions in wide orbits around \\the Barnard's Star from CanariCam mid-infrared imaging}
%
\maketitle

\begin{abstract}
   We have performed mid-infrared imaging of Barnard's Star, one of the nearest 
   stars to the Sun, using CanariCam on the 10.4~m Gran Telescopio Canarias.
   We aim to investigate an area within 1--10 arcsec separations, 
   which for the 1.83\,pc distance of the star translates to 
   projected orbital separations of 1.8--18\,au ($P >$ 12\,yr), which have not been 
   explored yet with astrometry or radial velocity programs.
   It is therefore an opportunity to enter the domain of distances where most 
   giant planets are expected to form.
   We performed deep imaging in the $N$-band window (Si-2 filter, 8.7\,$\mu$m) reaching a 3$\sigma$ 
   detection limit of 0.85\,$\pm$\,0.18~mJy and angular resolution of 0.24 arcsec, close to the 
   diffraction limit of the telescope at this wavelength. A total of 80 min on-source 
   integration time data were collected and combined for the deepest image.
   We achieved a dynamical range of $8.0\pm0.1$\,mag in the 8.7 $\mu$m band, at angular separations from 
   $\sim$2 to 10 arcsec and of $\sim$6--8\,mag at 1--2\,arcsec. No additional 
   sources were found. Our 
   detectability limits provide further constraints to the presence of 
   substellar companions of the Barnard's Star. According to solar metallicity 
   evolutionary models, we can exclude companions of masses larger than 
   15~$M_{\rm Jup}$ ($T_{\rm eff}>400$\,K), ages of a few Gyr, and located 
   in $\sim$3.6--18\,au orbits with a $3\sigma$ confidence level. This minimum 
   mass is approximately 5 $M_{\rm Jup}$ smaller than any previous imaging survey that 
   explored the surroundings of Barnard's Star could restrict.
   \end{abstract}


\section{Introduction}
   Optical and near-infrared high-contrast observations are of 
   major importance in the study of substellar objects. Direct 
   imaging searches for brown dwarfs and planets around stars 
   explore a range of physical separations complementary to 
   that of radial velocity or transit methods and provide key 
   information on their formation processes. Benefit of 
   detecting the direct light of planets is that it enables 
   their extensive characterization. We can determine their 
   ages, masses, radii, effective temperature and, in particular, 
   through spectroscopy we can have insight into their complex 
   atmospheres \citep{2012ApJ...754..135M, 2013ApJ...776...15C}. Yet, reaching a sufficiently 
   high-luminosity contrast at arcsecond level angular 
   separations from the target star is still very challenging. 
   As a result, in comparison with the transit method or 
   precision radial velocity measurements, direct imaging 
   brought only a few discoveries so far, e.g.,
   \citet{2004A&A...425L..29C}, \citet{2008ApJ...673L.185B}, 
   \citet{2008Sci...322.1348M, 2010Natur.468.1080M}, 
   \citet{2010Sci...329...57L}, \citet{2013ApJ...772L..15R}, 
   \citet{2013ApJ...763L..32C} and \citet{2015ApJ...804...96G}.   
   Most of the planets found by imaging are massive 
   ($M>5-10~\rm M_{Jup}$), young ($\tau <$ 500\,Myr), 
   and are located at relatively large projected physical 
   separations ($\sim$10--30 to 1000\,au) around their 
   host stars. Thus, each single discovery extending the known 
   population of imaged planets has still an important impact 
   on the field.
\begin{table*}
\caption{Observing log of GJ~699 with CanariCam at the GTC}             
\label{tab1}      
\centering          
\begin{tabular}{c c c c c c c c c c}     
\hline\hline       
OB & Observation & Start time & Savesets per & Nod    & Saveset  & On-source & Instrument & PWV  & Readout \\
   & date        &  (UTC)     & nod beam     & cycles & time (s) & time (s)  & PA ($^\circ$)   & (mm) & mode  \\  
  \hline  
   1 & 2012 July 29 & 23:13:18.7 & 8  & 10 & 6   & 3$\times$403.7 & 0   & 8.6--9.3 & S1R1\_CR  \\
   2 & 2012 July 30 & 00:20:01.6 & 8  & 10 & 6   & 3$\times$403.7 & 90  & 8.6--9.3 & S1R1\_CR \\
   3 & 2013 June 09 & 03:40:42.2 & 29 & 12 & 1.5 & 3$\times$431.9 & 0   & 6.7     & S1R3    \\
   4 & 2013 June 10 & 04:18:36.2 & 29 & 10 & 1.5 & 3$\times$359.9 & 300 & 6.7     & S1R3    \\
\hline                  
\end{tabular}
\end{table*}   

   In the formation process, gravitational collapse energy is 
   released and heats the interior of a planet but, as there is no 
   internal source of energy, planet cools and fades down with age.
   Giant planets at ages younger than 500\,Myr are more easily detectable 
   since at that stage their self-luminosity is still significant. 
   However, the vast majority of stars from the solar vicinity 
   ($d\sim$2--10\,pc) are relatively old, with ages similar to the age of 
   the Sun. Any potential planet from a nearby system, unless 
   orbiting at close separation from its host star, where it is 
   strongly irradiated, is expected to have cooled down to effective 
   temperature below 500--600\,K. Its spectral energy distribution
   would peak in the mid-IR range. Moreover, a solar-type star will 
   be relatively faint at these wavelengths, so that the contrast 
   necessary to detect a planet or a brown dwarf companion will be lower. 
   Currently, direct imaging of planets is feasible with the use 
   of {\em Hubble Space Telescope (HST)} or the largest ground-based facilities 
   equipped with adaptive optics systems, e.g. Gemini/NICI, Keck/NIRC2, 
   VLT/NaCo, Subaru/HiCIAO, operating in the optical 
   or near-IR regime. Searches require proper observing and data 
   processing techniques which attempt to remove diffracted light 
   like Lyot coronography \citep{2001ApJ...552..397S} or nulling 
   interferometry \citep{1998Natur.395..251H, 2000SPIE.4006..328S} 
   and to suppress the speckle background \citep[Angular Differential Imaging;][]{ 
   2006ApJ...641..556M, 2007ApJ...660..770L}. In 
   this work, we explore the potential of ground-based imaging at 
   mid-IR wavelengths to directly detect planetary-mass objects.
   
   Barnard's Star (GJ\,699, V2500\,Oph) is the fourth-closest individual 
   star and the second-closest system to the Sun currently known, after 
   the triple $\alpha$\,Centauri stellar system. It is also the closest 
   star in the Northern hemisphere. It was classified as an M4.0V-type 
   red dwarf \citep{1995AJ....110.1838R}, located at 1.824$\pm$0.005~pc 
   \citep{2007AA...474..653V} and moving at the highest known proper 
   motion $\mu=10.37$ arcsec per year. 
   A slow rotation with a period of about 130~d \citep{1998AJ....116..429B}, 
   low magnetic activity \citep{1999A&AS..135..319H} and other age indicators suggest that it  
   is probably older than the Sun.
   Since its large proper motion was measured by \citet{1916AJ.....29..181B}, the star received 
   much attention from astronomers. Its bolometric luminosity and effective 
   temperature were found to be $\left(3.46\pm0.17\right)\times10^{-3}$~L$_{\odot}$
   and $3134\pm102$~K \citep{2004AJ....127.2909D}. From mass--luminosity 
   relations for very low mass stars, \citet{2000A&A...364..217D} 
   determined the mass of GJ\,699 to be 0.158$\pm$0.013~M$_{\odot}$.
   Particularly, extensive studies have been carried out refining the limits 
   of possible planetary-mass companions \citep[e.g.,][]{1999AJ....118.1086B, 
   2012AJ....144...64D, 2013ApJ...764..131C}. Here, we 
   present the results of mid-IR imaging of this star using CanariCam at 
   the GTC. The following sections of the paper describe the performed 
   observations, reduction and analysis of the collected data, determination 
   of the sensitivity limits, its translation to the physical parameters (mass, 
   separations) of detectable companions and comparison with the previous surveys.

\begin{figure}
 \centering
 \includegraphics[scale=0.7]{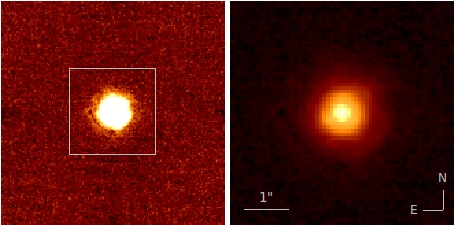}
 \caption{Final image of the Barnard's Star at 8.7 $\mu$m obtained with 
 CanariCam at the GTC. The used on-source time is 80 min. Field of view shown 
 in the left-hand image is 13\,$\times$\,13 arcsec. Right-hand image is the enlarged 
 5\,$\times$\,5 arcsec central part with a different contrast scale, marked with 
 a grey square on the larger field image. North is up, east to the left.}
 \label{ccimg}
\end{figure}

%
\section{CanariCam observations}
   Observations of the Barnard's Star were carried out in queue mode, 
   during the nights of 2012 July 29 and 2013 June 9 and 10 (UT). We 
   used the mid-infrared camera CanariCam \citep{2008SPIE.7014E..0RT}
   operating at the Nasmyth-A focal station of the 10.4~m Gran Telescopio 
   Canarias (GTC) at the Roque de los Muchachos Observatory on the island 
   of La Palma. CanariCam was designed to reach the diffraction limit of 
   the GTC at mid-IR wavelengths (7.5--25~$\mu$m). The instrument uses a 
   Raytheon 320$\times$240 Si:As detector with a pixel scale of 
   $79.8\pm0.2$\,mas, which covers a field of view of $25.6\times19.2$ 
   arcsec on the sky. We imaged our target in the 10 micron window, using 
   a medium-band silicate filter centred at $\lambda=8.7~\mu \rm m$ 
   ($\Delta\lambda=1.1~\mu \rm m$). The choice of this particular 
   bandpass was a compromise between the instrument performance, in 
   particular the filters transmissivity, and the sky background 
   contribution significantly higher at the $N$ broad-band and other 
   narrow-band filters than at the Si-2 filter. Si-2 is also favoured 
   by a better spatial resolution, since the diffraction disc is larger 
   at the available narrow-band filters at longer wavelengths.

 \begin{figure*}
 \centering
 \includegraphics[scale=0.79]{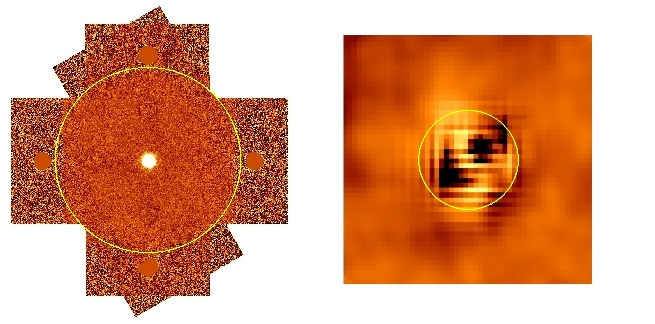}
 \caption{Left: image mosaic of the Barnard's Star, 
 processed and stacked to encompass the full area available in the 
 CanariCam data. Right: {\it WISE W3} image of the Barnard's Star, 
 after subtraction of the PSF of the target. The yellow circle on both 
 images marks an area within 14 arcsec radius around the centroid position 
 of the star. The CanariCam mosaic image fully covers the area up to 
 this separation and extends up to 20 arcsec
 but with incomplete coverage and lower sensitivity.
 The shown field of view of the {\it WISE} image cut-off is 70$\times$70 arcsec.
 No additional source is detected within this separation.
 Residuals of the star after PSF subtraction extend to roughly 14 arcsec.
 North is up and east is to the left.}
 \label{mosaic}  
\end{figure*}    
   
   Observations were performed with the standard chopping and nodding 
   technique used in the mid-IR to remove the sky emission and radiative 
   offset. Chopping consists of switching the telescope secondary mirror 
   at a typical frequency of a few (2--5) Hz between the position of 
   the source (on-source) and the nearby sky (off-source). This rapid 
   movement of the secondary mirror allows subtraction of the sky 
   background emission that is varying in time at frequencies below the 
   chop frequency. Movement of the secondary mirror changes the optical 
   configuration of the telescope, resulting in two different emission 
   patterns seen by the camera and producing a spurious signal termed 
   the radiative offset in the chop-differenced images. To remove the 
   radiative offset, the telescope is moved between two nod positions to
   swap over on- and off-source positions.
   We used a ABBA nodding sequence and `on-chip' chopping and nodding, 
   with a chop-throw and nod offset of 8 arcsec, a chopping frequency of 
   1.93--2.01~Hz and a nod settle time of about 45\,s. On chip method is recommended 
   whenever the scientific target is point-like, since both on-source and 
   off-source chop positions contain the signal of the target inside the 
   detector field of view and can be aligned and combined. On-source 
   integration time in each of the four observing blocks (OBs) was 20 min,
   divided into three data files composed of a set of images (savesets) at 
   subsequent chopping and nodding positions. Individual frames of 26 and 
   19\,ms exposures were co-added by CanariCam control software to savesets 
   of 1.6 and 6\,s for S1R3 and S1R1\_CR readout modes, respectively.
   A detailed observing log is given in Table \ref{tab1}.

   In total, we integrated the source for 80 min. Sky conditions during 
   the observations were photometric. Precipitable water vapour (PWV) as 
   measured by the Instituto de Astrof\'isica de Canarias real-time PWV monitor, was 8.6--9.3 and 
   6.7\,mm for OB 1, 2 and OB 3, 4, respectively. We observed with a set 
   of three different position angles of the instrument on the sky, 0$^\circ$, 
   90$^\circ$ and 300$^\circ$, and with two chop position angles of the secondary 
   mirror, 0$^\circ$ and 90$^\circ$, to guarantee that the chop offset was 
   done along the $X$-axis. By that, we wanted to avoid the loss of 
   potential objects in the regions overlapping with the negative images 
   of the star from the off-source chop positions or in parts obscured 
   by the horizontal cross-talk features that appear for bright sources \citep{2003SPIE.4841..169O}. 
   Also, we could discard eventual contaminants along the chop axis. 

\section{Data reduction and analysis}
\subsection{Image processing}
   The data were processed using standard routines within the 
   {\sc iraf\footnote{{\sc iraf} is distributed by the National Optical 
   Astronomy Observatories, which are operated by the Association 
   of Universities for Research in Astronomy, Inc., under 
   cooperative agreement with the National Science Foundation.}} 
   environment. 
   CanariCam images are stored in the standard multi-extension fits 
   files, with a structure of [320,\,240,\,2,\,M][N], where 320 and 240 
   are the image pixel dimensions, 2 is the number of chop positions,
   M of savesets and N of nod positions. 

   Off-source savesets, where the position of the secondary mirror is not 
   aligned with the primary mirror, were subtracted from the corresponding 
   on-source savesets, for respective nod beam position. These chop-/sky-subtracted 
   frames, where the star is located at the centre of the detector, 
   were then aligned to correct for very small misalignments (of less than 3 pixels), 
   and each pair corresponding to the A and B nod positions were combined, to subtract 
   the radiative offset. The sky-subtracted frames where multiplied by $-$1 to recover 
   the negative contributions of the star (off-source position of the secondary mirror). 
   Because the negatives in the A and B nod positions do not overlap, being at opposite 
   sides and at 8 arcsec of the on-source central location, they were radiative-offset 
   subtracted before they were aligned. Residual detector levels constant along single 
   columns or lines but varying across these remained in both the positive and negative 
   chop- and nod-subtracted frames; these were background fitted (masking the target) and
   subtracted. The alignment itself was applied at once, to all (positive and negative) 
   images of consecutive repetitions of an OB, relative to a same reference image and so that the 
   target has its centroid located on an integer pixel position.  
   Before aligning, the images were copied into larger ones to avoid the trimming of outer 
   data regions.  
   Then the frames were average-combined per repetition or altogether, 
   using a shallow sigma upper and lower clipping to discard occasional short transients 
   and extreme pixel values. Each combination involved masking the negative counts of the 
   target. OB 4 had cross-talk features that we had removed. OB 2 and 4 were acquired 
   with position angles differing from the North-up East-left orientation. For OB 2, 
   we transposed the stack to North-up East-left, whereas for OB 4, we resampled 
   the stack. In total, the data of OB 1, 2 and 3 were thus resampled only once, whereas 
   those of OB 4 twice. For the combination of the stacks of the different OBs, the 
   stacks were flux-scaled according to their zero-point magnitude -- as measured on the
   target -- and weighted inversely proportional to the scaled variance of their 
   background noise and the square of the full-width half-maximum (FWHM) of the target. 
   The central part of the final reduced image, with total on-source time of 
   80~min is displayed in Fig.~\ref{ccimg}. 
   An image mosaic of the whole covered area is displayed in Fig.~\ref{mosaic}. 
   This all-epoch mosaic extends to distance of about 20 arcsec (37 au) to the target 
   at the four cardinal directions. 
\begin{figure}
 \centering
 \includegraphics[scale=0.79]{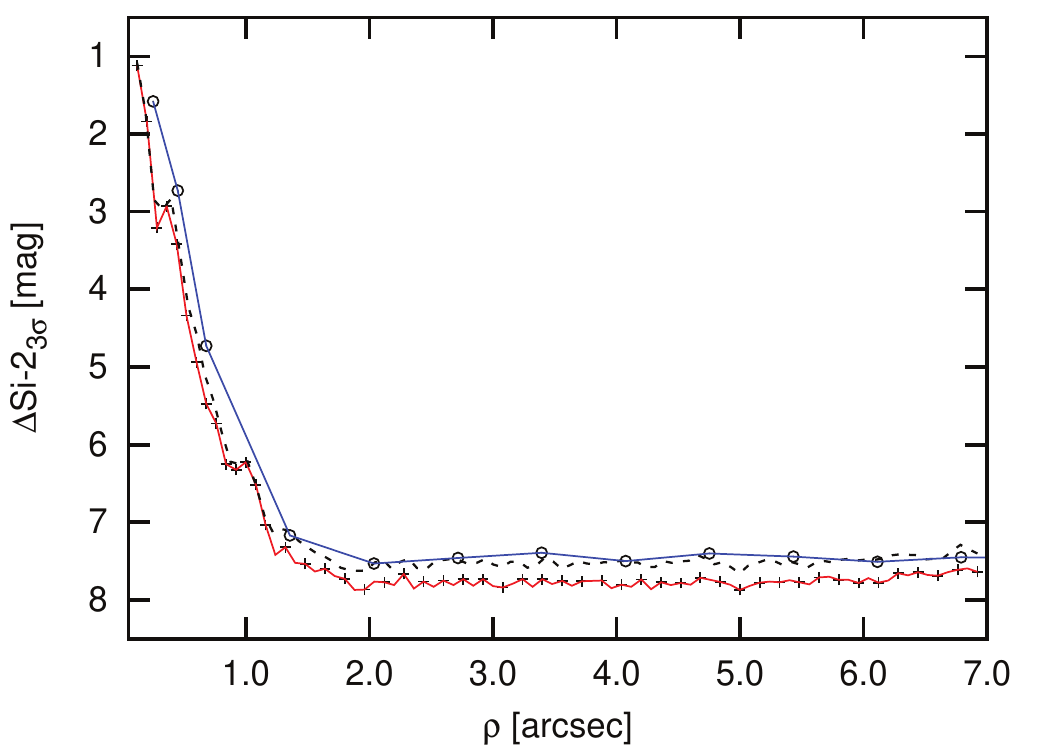}
 \caption{The Si-2 band (8.7 $\mu$m) $3\sigma$ contrast curves, derived on the deepest obtained image
 from artificial source injection (blue line with circle marks) and from 
 background noise computed in concentric annuli around the star (red 
 line with cross marks). We reach $\Delta$Si-2\,$\sim8$\,mag 
 at separations larger than 2 arcsec. The dashed line shows the sensitivity
 of combined single epoch images OB\,3+4, which reaches a detection limit 
 of about 0.25\,mag less beyond 2 arcsec separation.
 The two bumps at about 0.3 and 0.8\,arcsec are caused by the presence 
 of first and second Airy rings.}
 \label{detectability}
\end{figure}
   \subsection{Sensitivity and detection limits}
   The star point-spread function (PSF) on the final obtained images has 
   an FWHM of 3.01 pixels corresponding to 0.24 
   arcsec, for the image including all-epoch data, and 2.89 pixels (0.23 arcsec) 
   for the image including single epoch data from OB 3 and 4. 
   It is close to the theoretical FWHM of the diffraction-limited 
   PSF, which for GTC is 0.19 arcsec at 8.7\,$\mu$m. 

   To determine the Barnard's Star magnitude in the Si-2 filter we used 
   the $J, H, K$ photometry from the Two Micron All Sky Survey 
   (2MASS; \citealt{2006AJ....131.1163S}) and $W1, W2, W3, W4$ from the 
   {\it Wide-field Infrared Survey Explorer} ({\it WISE}) All-Sky and AllWISE Source 
   Catalogs (\citealt{2010AJ....140.1868W}).
   We converted the 2MASS and {\it WISE} magnitudes into fluxes using the
   corresponding Vega zero points and interpolated the values to obtain 
   average flux at 8.7\,$\mu$m through a least-squares fit to a power 
   function. {\it WISE} $W2$ measurement was not used in the fit because 
   of saturation, affecting sources brighter than approximately 6.7\,mag 
   in this band. The calculated Si-2 brightness of Barnard's Star is 
   $4.12\pm0.19$\,mag, using the Vega system zero point determined for 
   this CanariCam filter. This value is very close to the $W$3 magnitude 
   of the star ($W3=4.036\pm0.016$\,mag). For our deepest image, that is, 
   the one combined from all OBs, the limiting 
   magnitude is $11.92\pm0.25$\,mag (0.85\,$\pm$\,0.18~mJy), estimated 
   using the ratio of the peak counts of the star to 3$\sigma$ background 
   noise.
 \begin{figure}
 \centering
 \vspace*{1mm}
 \includegraphics[scale=0.56]{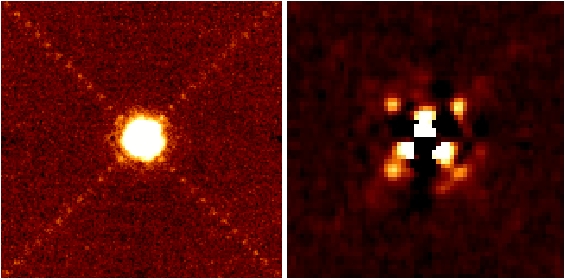}
 \caption{Same as Fig.~1, but with artificial sources inserted along the 
 diagonals in left-hand panel, and after PSF subtraction and the addition 
 of artificial sources in right-hand panel.} 
 \label{ccimg_addstars}
\end{figure}   

   In Fig.~\ref{detectability}, we present the 3$\sigma$ contrast curves for CanariCam 
   image of Barnard's Star. The two solid lines shown in the plot correspond to two 
   different contrast calculation methods that have been used. In the first 
   approach, we computed the background noise, $\sigma$, as a function of 
   radial separation from the target star, by measuring the standard deviation 
   in a 1 pixel wide concentric annuli around the star. The 3$\sigma$ noise 
   counts were converted to contrast (delta magnitudes between the primary 
   star and the measured quantity, noise in this case) by relating to the 
   peak pixel value of the star PSF. The contrast curve obtained using this 
   method is plotted in Fig.~\ref{detectability} with a red solid line with cross marks.
    
   In the second method, we used artificial sources to estimate the detection 
   limits and contrast as a function of radius. Artificial objects were produced 
   using the PSF model of the primary and inserted on the diagonal axes, with 
   intervals of $\sim 2\times$FWHM ($\sim0.6$\,arcsec) and decreasing steps
   in the inner parts of the image. We considered as a detection when the added
   object at a given magnitude and radial distance is detected through visual
   inspection in at least three of the four diagonal positions. A simulated source
   with its PSF retaining characteristic stellar shape and visible marginally above
   the background noise corresponds typically to a signal to noise of 3--5.
   To examine the inner area (within 1 arcsec) around the star, we performed 
   target PSF subtraction on the images with inserted artificial sources by 
   rotation or flipping along an axis, by previously measuring the centre 
   and semi-major axis of the ellipse isophote closest to the level of the 
   candidates ($\sim$10$\sigma$) we expect to find in the PSF wings.
   Example images with simulated artificial objects are shown in Fig.~\ref{ccimg_addstars}.
   Results from this two methods, which were found consistent, yield that we 
   reach a dynamical range in Si-2 of about 6\,mag at $\sim1$\,arcsec 
   separation from the target star and a maximum of 8\,mag at separations 
   $\ga2$\,arcsec. The second method systematically yields detection limit
   as a function of projected separation from the Barnard's Star that is 
   about 0.25 mag brighter than obtained with the first method.

\begin{figure*}
 \centering
 \includegraphics[scale=0.94]{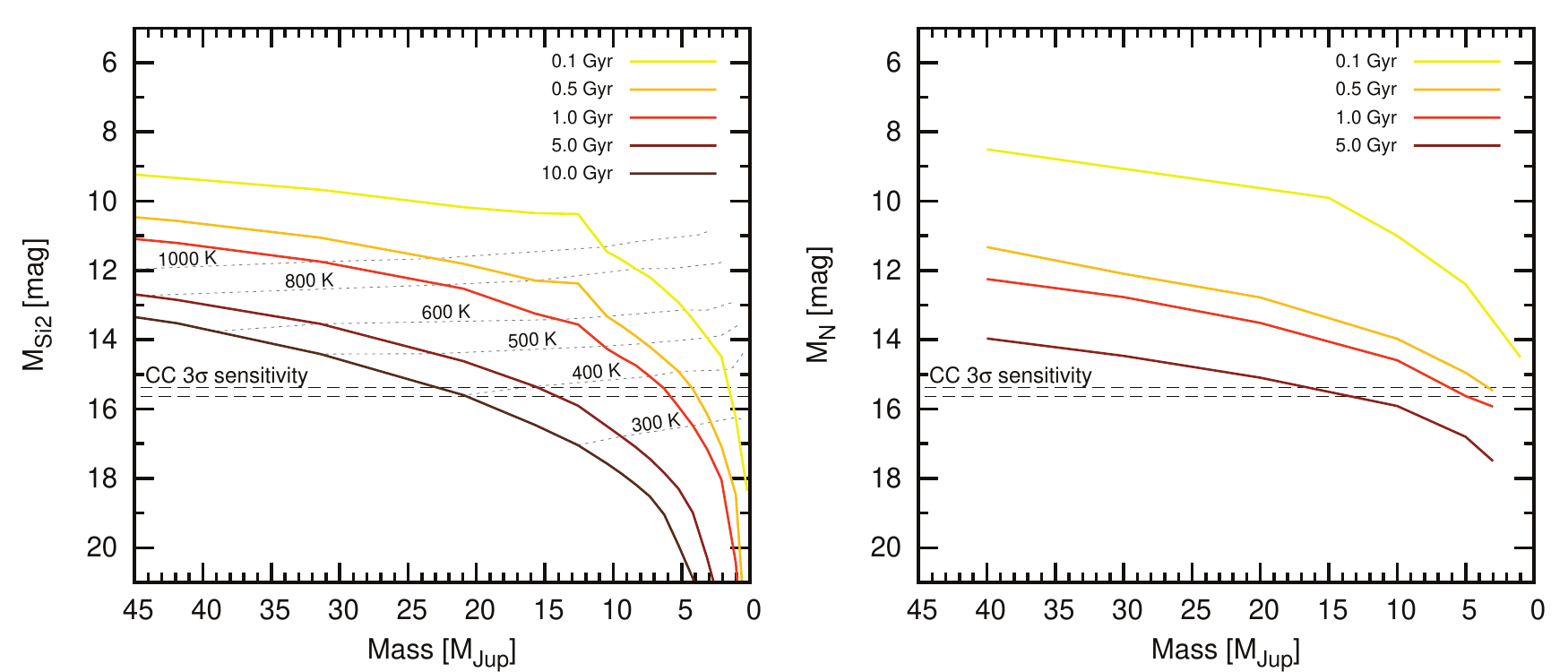}
 \caption{Theoretical absolute magnitudes versus mass of brown dwarfs and 
 giant planets at Si-2 8.7\,$\mu$m (left) and $N$ bands (right). For 
 Si-2 we used the Ames-COND models at ages of 0.1, 0.5, 1, 5 and 10 Gyr and 
 for $N$-band we used models by \citet{1997ApJ...491..856B}. Dashed 
 horizontal lines mark our 3$\sigma$ detection limit range from single epoch 
 observations: M$_{\rm Si-2}$\,=\,$15.36\pm0.28$ and from the deepest 
 image obtained using both epochs: M$_{\rm Si-2}$\,=\,$15.62\pm0.25$\,mag.
 Also several isotherms are plotted in the left-hand panel with grey dotted lines.
 According to the models, this implies that companions more massive 
 than $\sim$15\,$M_{\rm Jup}$ or with T$_{\rm eff}>400$\,K at solar 
 ages would have been detected.}
 \label{models}
\end{figure*}   
   
   At the range of projected orbital separations explored in this search, 
   with roughly one year period separating the two observations, the orbital 
   motion of a potential companion may not be negligible. We estimate that 
   a planetary mass companion at a 18 au circular orbit orientated face-on 
   would move $\sim$0.3 arcsec over one year baseline which is approximately 
   the FWHM of our CanariCam images, and up to 0.8 arcsec at 3 au orbit (1.5 arcsec angular separation). At 
   an edge-on orientation, the shift could be higher than 0.3 arcsec at 
   separations closer than 5 au. Since for some range of parameter space the 
   displacement would be higher than the spatial resolution of the images, 
   we provide also the detection limit of single epoch data.
   The dashed line in Fig.~\ref{detectability} is the contrast curve 
   determined on the image obtained using only first epoch images 
   (OBs 3 and 4, Table \ref{tab1}). The achieved detection limit in 
   this case, with a total of 40 min on-source time, is about 0.25 mag 
   lower than that achieved with the use of both epochs observations.

\section{Constraints on the presence of companions}
\subsection{Physical interpretation of detection limits}
   In the following we relate the detection limits of our search to 
   the physical properties of brown dwarf and planetary companions, 
   specifically to their masses, effective temperatures and luminosities. 
   A complication inherent to objects in the substellar domain is the 
   continuous cooling in the course of their evolution. It precludes 
   the possibility to estimate the mass applying unique relations 
   independent of age, such as the mass--luminosity relation for the 
   main-sequence stars. In this case, we need to rely on theoretical 
   models providing a grid of luminosities, temperatures, synthetic 
   photometry as a function of masses and ages. In this work, we used 
   the Ames-COND models (\citealt{2001ApJ...556..357A}; \citealt{
   2003A&A...402..701B}; \citealt{2012RSPTA.370.2765A}) for solar 
   metallicity and the models of giant planets and brown dwarfs by 
   \citet{1997ApJ...491..856B}. 
   Both the COND and the Burrows et al. (1997) models apply to $T_{\rm eff}<1300$\,K 
   and extend down to 100 K. They include the formation of dust in the atmospheres 
   of this objects, however they neglect its opacity, considering that the dust 
   grains settle below the photosphere.  
   To compute the synthetic magnitudes 
   for the Si-2 8.7\,$\mu$m band we used the Phoenix Star, Brown Dwarf 
   \& Planet Simulator available online\footnote[2]{http://phoenix.ens-lyon.fr/simulator/index.faces}. 
   We input the transmission 
   file of the Si-2 filter and obtained the isochrones for a set of 
   five different ages: 0.1, 0.5, 1, 5 and 10 Gyr. 

   At the relatively old ages ($>$1 Gyr), substellar companions in 
   the planetary-mass regime have cooled down to temperatures below 
   600\,K, according to theoretical evolutionary models 
   \citep{1997ApJ...491..856B, 2000ApJ...542..464C}. 
   At these temperatures, brown dwarfs and giant planets are very 
   faint in the near-IR and emit most of their flux in the 
   mid-IR. As an example, while a 5\,$M_{\rm Jup}$ planet near-IR 
   flux decrease from $M_{\rm J} \sim 15$\,mag at 10\,Myr to
   $M_{\rm J} \sim 25$\,mag at 4.5\,Gyr, the mid-IR emission only 
   changes from $M_{\rm N}$\,$\sim$\,11.5\,mag to $M_{\rm N}$\,$\sim$\,16
   mag in the same interval of age.

   In the left-hand panel of Fig.~\ref{models}, we plot the Si-2 absolute 
   magnitudes versus masses produced by the models, in comparison 
   with the 3$\sigma$ detection limit of our CanariCam observations. 
   For ages of 1 Gyr and younger, objects down to 5\,$M_{\rm Jup}$ 
   would have been detected. At solar ages, we are sensitive to objects 
   with masses higher than approximately 15~$M_{\rm Jup}$ or with 
   effective temperatures above 400\,K and at projected separations 
   from $\sim$3 to 18\,au. At smaller separations, around 1\,arcsec 
   ($\sim$2\,au), we could have detected objects with masses above 
   $\sim$20\,$M_{\rm Jup}$. 
   
   Several indicators point that the Barnard's Star belongs to the
   older population. The non-detection of lithium in the atmosphere
   imposes a minimum age of 20\,Myr \citep{2004ARA&A..42..685Z}. The level of 
   chromospheric activity suggest an age above 600--800\,Myr \citep{2006PASP..118..617R}.
   It has a very low X-ray luminosity ($\log{L_x}=26$) indicating 
   low level magnetic activity \citep{1999A&AS..135..319H, 
   1981ApJ...245..163V}. \citet{1996AJ....111..466E} estimated 
   the age of GJ\,699 at 10~Gyr based on the Ca~{\sc ii} index. Also its 
   high space velocities \citep{1992ApJS...82..351L, 
   1996AJ....111..466E}, lower than solar metallicity 
   [Fe/H]=$-0.39\pm0.17$ \citep{1997AJ....113..806G} and a 
   probable long rotation period of approximately 130 days 
   \citep{1998AJ....116..429B} are all consistent with a 
   relatively advanced age of 7--12~Gyr.
   At that age, our sensitivity limit enables us to detect 
   companions more massive than approximately 20\,$M_{\rm Jup}$ 
   with $T_{\rm eff}>450$\,K. 

   For a comparison, we have checked the models 
   by \citet[right-hand panel of Fig.~\ref{models}]{1997ApJ...491..856B}
   that provide synthetic photometry in the 
   {\it N} band (10 $\mu \rm m$) at ages of 0.1, 0.5, 1 and 5 Gyr for 
   objects with masses below 40\,$M_{\rm Jup}$. We find that the 
   two models are in a fairly good agreement and give a similar 
   mass and $T_{\rm eff}$ constraint using the same Si-2 
   sensitivity limit. We interpret that the slightly brighter 
   magnitudes for the same masses yield by the $N$-band 
   isochrones result from a wider bandpass and longer central 
   wavelength of this filter, at which the flux of very cool 
   objects is expected to be larger, as well as from differences 
   between the two models.
   
\subsection{Comparison with previous searches and \textbf{\textit{WISE}} data}
\begin{figure}
\centering
\includegraphics[scale=0.83]{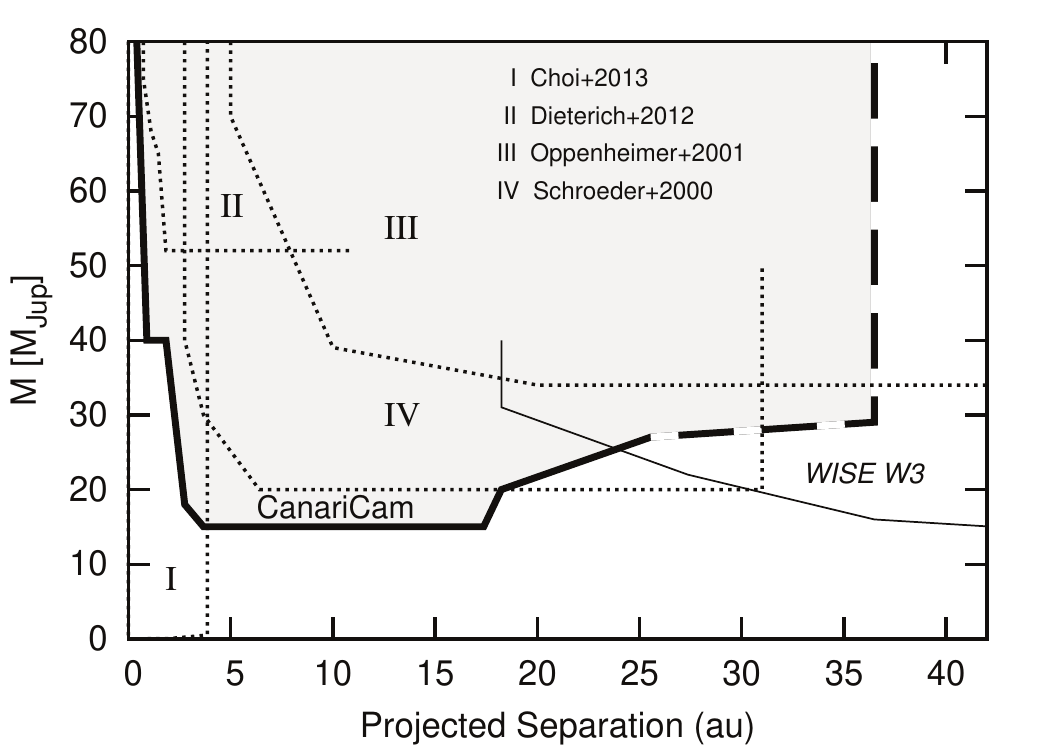}
\caption{Constraints on the presence of substellar companions around 
the Barnard's Star. Comparison of obtained CanariCam detection limits 
(thick line, shaded area) with results from previous searches and limits 
determined using {\it WISE W3} image. Corresponding surveys marked in the 
diagram are described in Section 4.2. The CanariCam area with dashed line 
border corresponds to the region at angular separations $>$14 arcsec, 
where the mosaic coverage is not complete.  }
\label{comparison}
\end{figure}

   First claim of planetary companions of Barnard's Star were 
   reported in the late 60s by \citet{1969AJ.....74..757V}. 
   Using astrometry, he detected perturbations in the proper 
   motion of the star consistent with two planets comparable 
   in mass with Jupiter. These planets however were definitively 
   ruled out by the radial velocity studies. 
   \citet{2013ApJ...764..131C} used Doppler measurements 
   obtained from Lick and Keck Observatories with a precision 
   of 2 m\,s$^{-1}$ during an 8 yr monitoring to preclude 
   planetary companions of masses above 2 Earth masses with 
   periods below 10 days and above 10 Earth masses with periods 
   up to 2 yr, save for face-on orbits. (area labelled `I' 
   in Fig.~\ref{comparison}). This result is supported by the 
   probability of 94.6--98.2\% that the orbital inclination 
   of putative companions is larger than 11$^\circ$--19$^\circ$.
   Various ground- and space-based imaging surveys dedicated to 
   detect ultracool companions targeted the Barnard's Star along 
   the past two decades. In this section, we focus on the ones 
   that reached the highest sensitivities up to date and compare 
   them with our results.  

   \citet{2012AJ....144...64D} carried out a large, volume-limited
   search for substellar companions to stars (mainly around M dwarfs) 
   in the solar neighbourhood within $\sim$10 pc. The survey was 
   performed with the {\it HST} NICMOS 
   instrument, using four filters centred at 1.10, 1.80, 2.07 and 
   2.22\,$\mu$m, for snapshot high-resolution imaging of 255 
   individual stars. Physical separations accessible in this search 
   correspond to mean semimajor axes between 5 and 70\,au. The minimum 
   detectable mass was estimated to be 42\,$M_{\rm Jup}$ at 3 Gyr or 
   52\,$M_{\rm Jup}$ at 5 Gyr. 
   The masses were derived using the \citet{2000ApJ...542..464C} models.   
   For the Barnard's Star however, as it 
   is one of the nearest stars from the sample, closer separations 
   were probed. For the survey limits on this particular star, we used 
   the information from the table 2 in \citet{2012AJ....144...64D}, 
   which gives the sensitivities achieved for each target at various 
   separations. The range of masses and separations of companions ruled 
   out by this survey is marked as area II in Fig.~\ref{comparison}.

   \citet{2001AJ....121.2189O} observed northern stars from the 8 pc 
   sample including GJ~699, in the optical (Gunn $r$ and $z$ filters) 
   and near-infrared wavelengths ($J, K$ filters). They used the Adaptive 
   Optics Coronograph instrument on the Palomar 1.5~m telescope for 
   the optical imaging and the Cassegrain Infrared Camera on the Palomar 
   5~m Hale Telescope, for the near-IR. For about 80\% of the surveyed 
   stars, companions more massive than 40\,$M_{\rm Jup}$ at 5 Gyr age 
   would have been detected at separations between 40 and 120 au. 
   The mass detection limits were determined using the \citet{1997ApJ...491..856B}
   evolutionary models.
   The area marked as III in Fig.~\ref{comparison} shows the limits of the lowest 
   mass detectable by the survey at the closest accessible separations, 
   given in tables 11 and 12 in \citet{2001AJ....121.2189O}.
%

   \citet{2000AJ....119..906S} searched for faint companions to 23 
   stars within 13 pc of the Sun using the {\it HST} Wide Field Planetary 
   Camera 2. Barnard's Star was imaged through $F675W$, $F814W$ 
   and $F1042M$ filters using short and long exposures ranging from 1.6 
   to 600\,s, to map different separations from the target star. 
   According to the limits shown in fig. 13 in the paper, their survey 
   was sensitive to objects of about $40~M_{\rm Jup}$ at angular 
   separations larger than 1.5\,arcsec, $30~M_{\rm Jup}$ at 
   $>2.0$\,arcsec and $20~M_{\rm Jup}$ beyond 3.5\,arcsec, for an age 
   of 5~Gyr. 
   The authors also employed the \citet{1997ApJ...491..856B} models to derive the limits of the survey
   in terms of companion masses.
   These limits are shown in Fig.~\ref{comparison} as area IV, 
   which extends up to $\sim$30 au.

   Fig.~\ref{comparison} summarizes in a simplified manner the detectability
   limits of the described surveys and of the CanariCam imaging search
   in terms of companion mass lower limits and projected physical 
   separations. 
   Additionally, we have included the detection limits of the search 
   for companions we have carried using the {\it WISE W}3 band data 
   at 12 $\mu$m. We have retrieved Barnard's Star's image from the {\it WISE} 
   All-Sky release using the IRSA Wise Image Service. To derive the 
   sensitivity in the inner part, we have previously performed a PSF 
   subtraction of the target using as a reference the PSF of a similar 
   brightness star located close to the Barnard's Star on the sky, after 
   scaling their peak fluxes. We then obtained the 3$\sigma$ limit of 
   sensitivity of $W3$ using the same method as for the CanariCam data 
   and compared to the Ames-COND models computed for this band. We 
   estimated a sensitivity of $W3$ = 9.7--9.8 mag at 14--15 arcsec 
   separations, 10.5 mag at 20 arcsec and a limit of 11.5 mag ($\sim$1.0 mJy) 
   at 40 arcsec and beyond. According to models, a (solar age) 
   companion of at least 15~$M_{\rm Jup}$ located at separation 
   $\geq$\,40\,arcsec ($\geq$ 73\,au) would have been detected. The obtained detection 
   limit in terms of sensitivity and of objects masses and effective 
   temperatures is in agreement with the general documentation of the 
   mission \citep{2010AJ....140.1868W} and other studies that employed 
   {\it WISE} data, e.g. \cite{2014ApJ...786L..18L}. This search for companions 
   using {\it WISE W}3 data which extends to wider separations is complementary 
   with our higher resolution CanariCam images and provides similar sensitivity. 
   
   Radial velocity and astrometry techniques exclude 
   planetary companions of less than 1~$M_{\rm Jup}$ with orbital 
   periods up to 2 yr \citep{2013ApJ...764..131C, 1999AJ....118.1086B}. 
   This detection limits are still far beyond the capabilities of any 
   other method, whereas at wider orbits direct imaging provides the 
   strongest constraints on the presence of companions. With CanariCam, 
   we could have detected companions more massive than $15~M_{\rm Jup}$ 
   at projected separations from $\sim$3 to 18 au for an age of 5~Gyr. 
   CanariCam observations have enabled us to set the strongest
   constraints on the presence of very low mass brown dwarfs in wide
   orbits (3.6--18\,au) with a 99\% confidence level. However, it
   should be noted that this statement is based on model evolution
   predictions. Were the models prove to be invalid for these least
   massive substellar objects, the obtained mass and temperature lower 
   limits would have to be reviewed. In particular, the Ames-COND 
   isochrones that were used, are available only for solar abundance. 
   Given the slightly sub-solar metallicity of the Barnard's Star, 
   this may result in a difference of the derived mass.

\section{Summary and final remarks}
   We performed a deep, high spatial resolution imaging of the Barnard's 
   Star at mid-IR Si-2 8.7~$\mu \rm m$ wavelength using CanariCam at the 
   10.4\,m GTC telescope. No companion candidates were found on the 
   obtained images. Our detectability limits provide further constraints 
   on the presence of substellar companions. With 80 min on-source 
   integration time we achieved sensitivity of 0.85\,$\pm$\,0.18~mJy 
   allowing to detect massive planets and brown dwarf companions down 
   to $15~M_{\rm Jup}$ which corresponds to effective temperatures above 
   400~K, assuming a solar age. Our search covers a field of 1--10\,arcsec 
   radius around the target star ($\sim$2--18 au for Barnard's Star 
   distance), which means that we can probe the domain of distances 
   where most giant planets are expected to form \citep{2014prpl.conf..619C}.

   This work demonstrates that the modern ground-based mid-IR imaging 
   instruments operating on 10-m class telescopes can reach angular 
   resolutions and sensitivity limits as good and, in certain cases 
   (e.g. nearby, relatively old stars) better than adaptive optics 
   systems in the optical or near-IR or space telescopes. This 
   technique presents a high potential to perform direct imaging 
   studies of brown dwarfs and exoplanets.
\newline
\newline
{\bf ACKNOWLEDGEMENTS}\vspace*{1mm}\newline
   We are grateful to the GTC staff for performing the CanariCam observations.
   This publication makes use of data products from the {\it Wide-field Infrared 
   Survey Explorer}, which is a joint project of the University of California, 
   Los Angeles, and the Jet Propulsion Laboratory/California Institute of 
   Technology, funded by the National Aeronautics and Space Administration. 
   Based on observations made with the Gran Telescopio Canarias (GTC), installed 
   in the Spanish Observatorio del Roque de los Muchachos of the Instituto de 
   Astrof\'isica de Canarias, in the island of La Palma.
   This work is partially funded by the national program AYA2010-20535 funded 
   by the Spanish ministry of Economy and Competitiveness (MINECO).

\bibliographystyle{mn2e}
\bibliography{ms}
\end{document}